# DIAMOnDS- Distributed Agents for MObile and Dynamic Services

M Aamir Shafi, Maria Riaz, Saad Kiani, Anjum Shehzad, Umer Farooq, Arshad Ali
*National University of Sciences and Technology, Rawalpindi 46000, Pakistan*
Iosif C Legrand, Harvey B Newman
*California Institute of Technology, Pasadena, CA 91125, USA*

Distributed Services Architecture with support for mobile agents between services, offer significantly improved communication and computational flexibility. The uses of agents allow execution of complex operations that involve large amounts of data to be processed effectively using distributed resources. The prototype system Distributed Agents for Mobile and Dynamic Services (DIAMOnDS), allows a service to send agents on its behalf, to other services, to perform data manipulation and processing. Agents have been implemented as mobile services that are discovered using the Jini Lookup mechanism and used by other services for task management and communication. Agents provide proxies for interaction with other services as well as specific GUI to monitor and control the agent activity. Thus agents acting on behalf of one service cooperate with other services to carry out a job, providing inter-operation of loosely coupled services in a semi-autonomous way. Remote file system access functionality has been incorporated by the agent framework and allows services to dynamically share and browse the file system resources of hosts, running the services. Generic database access functionality has been implemented in the mobile agent framework that allows performing complex data mining and processing operations efficiently in distributed system. A basic data searching agent is also implemented that performs a query based search in a file system. The testing of the framework was carried out on WAN by moving Connectivity Test agents between AgentStations in CERN, Switzerland and NUST, Pakistan.

## 1. INTRODUCTION

Distributed Services Architecture (DSA), conceptualized from [1], having support for mobile agents between services, offer significantly improved communication and computational flexibility. The support for mobile agents allows several potential advantages like asynchronous communication and disconnected operation, remote interaction and adaptability, parallel execution and autonomous mobility among loosely coupled services in the distributed environment. This prototype system is based on the above mentioned architecture and it provides a secure and flexible framework which can be used for communication and coupling of distributed services used in High Energy Physics (HEP) [1].

The prototype design consists of AgentStations, which provide the runtime environment for agents and dynamically establish a peer-to-peer relation among them using a discovery and registration service. The AgentStation is made available as a global network service using the Jini technology which can be enhanced with other protocols including JXTA. [3] or UDDI/WSDL [4] because of the use of generic interfaces. The mobile agent acting on behalf of its user can move to different AgentStations to autonomously perform its task. Mobile agents may provide their downloadable GUI's [5] to clients in their dynamically downloadable proxies, to control and monitor their activities. Basic security has been implemented to establish trust among AgentStations and to avoid the mobile agents doing any harm to the AgentStation or vice versa. The system design avoids single point of failure, uses the network bandwidth efficiently by using mobile agents and aims to offer reliable and secure support for large scale distributed applications in real environment.

The prototype is an extensible one, which means that a developer can extend the generic agent to make a customized agent (mobile service) and make it part of DIAMOnDS network. Three data access agents have been implemented to validate and make use of the prototype. These agents can move to remote AgentStations and provide its user access to the file system or databases provided by the remote agent station. The agents' downloadable GUI can be used by the end user for accessing data on remote sites.

### 1.1. Concept

Distributed Services Architecture (DSA) [1] allows distributed software components residing on the network to be published, discovered and invoked by each other. It also allows a software programmer to model programming problems in terms of distributed services offered by components to anyone, anywhere over the network.

From the official Jini architecture specification [2], Jini is defined as: "*A distributed system based on the idea of federating groups of users and the resources required by those users. The focus of the system is to make the network a more dynamic entity that better reflects the dynamic nature of the workgroup by enabling the ability to add and remove services flexibly*."

Jini services form communities, thus allowing them to register, discover and invoke one another. Jini services submit proxies to the Jini Registration Service called Lookup Service (LUS) which are downloaded to the client side on request and are responsible for handling all communication between the client and the service. The concept of leasing ensures that all the services have updated information about all other services, thus addressing the issue of reliability by providing fault tolerance. The support for code mobility is a unique feature of Jini over other Service Oriented Architecture implementations which has been crafted to build a mobile agents platform in this project.

Mobile agents help to make robust applications that are able to overcome network latency and reduce network bandwidth by sending an agent to remote location where it can execute locally, act autonomously and adapt dynamically to heterogeneous environments [6].

This paper highlights the main features of a new mobile agents' framework for dynamic services based on Jini [2]. Section 2 discusses how code mobility is achieved in DIAMOnDS. Section 3 explains the system architecture and the four main modules developed in the prototype system. Section 4 discusses the security mechanism implemented. Section 5 and 6 describe the specific agents built using the prototype system and the tests conducted to prove the validity of the system. After concluding in the Section 7, current status and the future work is discussed in Section 8.

## 2. CODE MOBILITY

Java/Jini provides support for code mobility that has been crafted to implement the mobility mechanism based on [8]. If an agent wants to move to another AgentStation (either from an AgentStarter or from an AgentStation), it marshalls itself and calls a method on that AgentStation's proxy which allows the marshalled object to move to that AgentStation. The proxy collects the marshalled object and transfers it to the AgentStation. On receiving the marshaled agent, the AgentStation unmarshalls that agent and after carrying out some security checks, it allows its execution if the checks are successful. If an agent is a 'Service' type (agents registered with LUS), then the AgentStation allows the agent to register with the LUS using the AgentStation's discovery and registration interface.

The agent is moved as a marshalled object from one AgentStation to another (and also from AgentStarter to AgentStation). This marshalled object has the state of an agent saved inside it but no class definition. To unmarshall the object, the receiving AgentStation must know its class definition. The agent carries with itself a URL to the location where its class file is located. This class file is packed in a signed jar file and placed in a downloadable folder of the HTTP Server. When an AgentStation unmarshalls the agent, it also loads its class files after downloading the jar file from the HTTP Server.

## 3. SYSTEM ARCHITECTURE

The services, both the agents and AgentStations in the DIAMOnDS framework ensure reliability by having the correct information by making use of Jini leasing mechanism. For example, if a DIAMOnDS AgentStation crashes unexpectedly, it will not have a chance to notify interested agents and other AgentStations of its unavailability in near future. At this point, Jini Lookup Service (LUS) will recognize that the lease for this AgentStation has not been updated in the last 'n' minutes (configured at deployment time) and will remove its registration. At the same time, the LUS will generate a notification about unavailability of the AgentStation to all the clients.

Similarly, if an AgentStation loses network connectivity, it will be unregistered from the Jini Lookup Service (LUS) automatically. The AgentStation service will render itself unregistered because it is unable to renew its lease by contacting a LUS and the LUS will un-register the AgentStation because it did not renew its lease. As soon as the network connectivity becomes available, the AgentStation will once again register with the LUS and renew its lease. During the time the network connectivity is unavailable, Jini service mechanism allows the station to carry out disconnected operation; waiting for its chance to join the Jini network and the system as a whole acts in a self healing manner requiring no manual interference by an administrator or a user.

Communication among DIAMOnDS services takes place through 'proxies', which are downloaded to the client machine without the need to have an administrator to install the code; all the required classes/interfaces needed by another service or client to communicate with the original service is downloaded at runtime. This feature allows building of dynamic downloadable proxies of DIAMOnDS services (AgentStation and agents) which can be tailored to meet the constraints of the environment on which the client module is executing. For example, a client module will have different constraints on a desktop machine and a PDA and the service communicating with the client may have to present different proxies to enable effective execution and reliable communication. Furthermore, these proxies are used by the client through an interface and the client need not be aware of the actual protocol being used by the proxy to communicate with the service. This encapsulation allows development of services for heterogeneous environments with little overhead and programming ease.

The above features employed in the DIAMOnDS framework allow a developer to extend a basic service from the DIAMOnDS framework and build new customized services with inherited fundamental DIAMOnDS features.

The agents in DIAMOnDS framework have been implemented as Jini services. This allows for agent discovery by remote clients, fault tolerance through leasing and Java Spaces [7] and remote monitoring of agents through the client module.

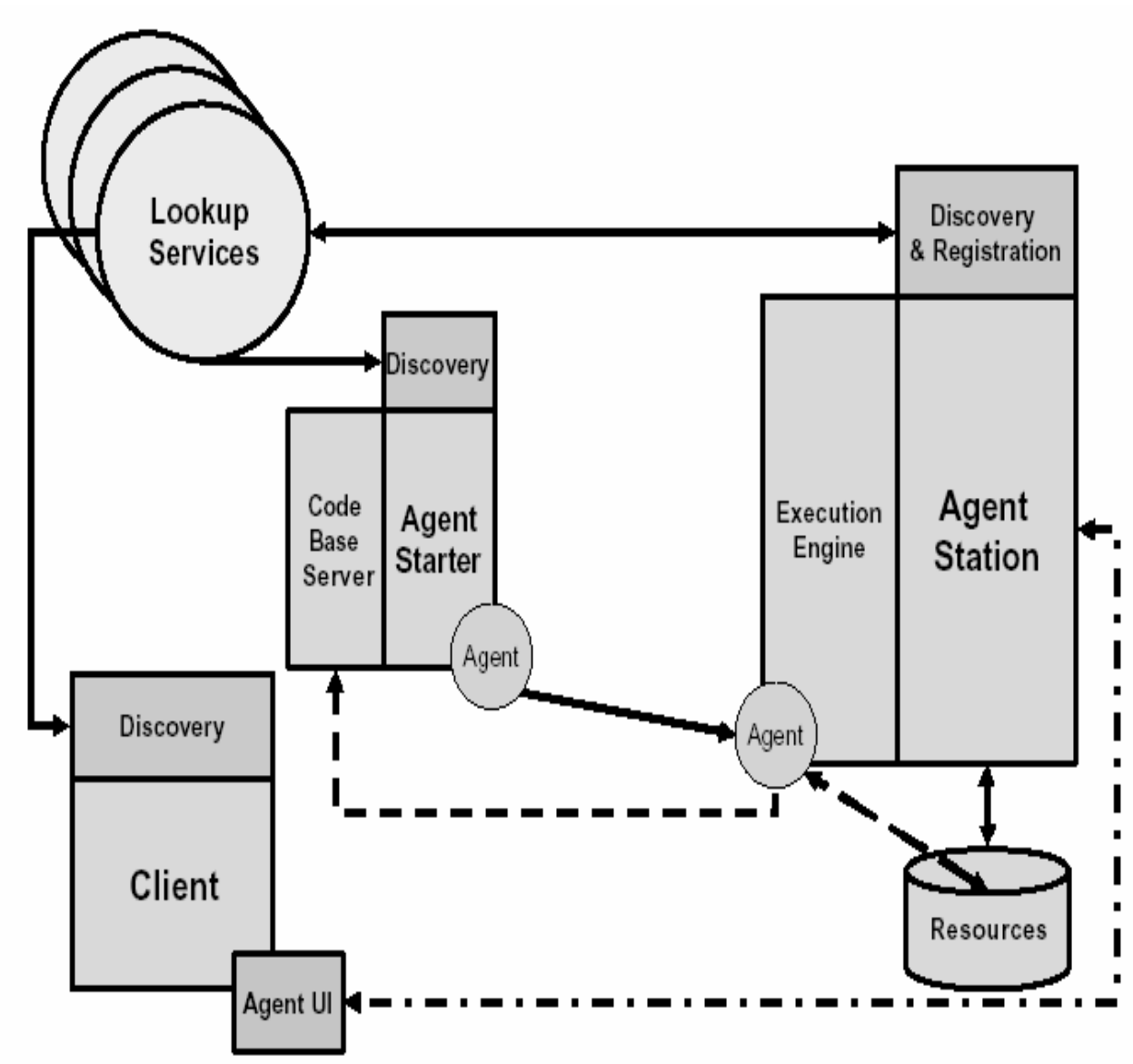

Figure 1: DIAMOnDS Architecture

Figure 1 illustrates the basic architecture of the DIAMOnDS framework. As mentioned previously, there are two services in this framework, the first is AgentStation which provides its services to mobile agents, and the second is the agent itself which provides its services to the end user. Both the services register themselves with the Lookup Services from where the clients, agents and the AgentStations discover them. The agent is loaded at the AgentStarter or the AgentStarter sub module in the AgentStation and discovers all the AgentStations through the Jini Discovery mechanism. The agent can then move to any AgentStation and get executed to perform any task assigned by the end user.

## 3.1. Agent Starter

AgentStarter module acts as a Jini client, thereby discovering all the AgentStations that are running at a particular time. The AgentStarter can load the agent from a jar file into the JVM either locally from the file system or remotely from a web server. As the agent is loaded into the JVM from the local file system, an HTTP server is started that serves the class file of the agent as and when required by other agent station or the clients.

The AgentStarter provides the functionality of loading an agent and then moving it to one of the discovered AgentStations.

## 3.2. Agent Station

AgentStation is the main component of this framework. AgentStations dynamically interact with each other and provide the platform on which an agent can be hosted and allowed to execute. Agents are authenticated by the AgentStations using the security mechanism of the framework before they are allowed to execute. Once an agent has been authenticated, it is allocated a separate thread having relevant priority.

Each AgentStation uses the Discovery & Registration service (a sub module in the AgentStation) to register itself with the Jini Lookup Services (LUSs) and also to maintain a list of all other registered services. This information is also provided to the hosted agents. On the request of a hosted agent, the AgentStation registers the agent with the Lookup Services. In effect, the agent uses the discovery and registration mechanism of the AgentStation to register it and to discover other services. The AgentStation has an Execution Engine which provides an execution environment to the authenticated agents that want to execute on this AgentStation. This execution engine can be thought of as a thread pool; each agent's execution is carried out in a separate thread. It allocates a new thread to every agent that requires execution and once the execution is complete, kills the thread.

Agent Starter sub module is also integrated into the AgentStation which provides services of loading an agent either from the local file system or from a remote HTTP server. Security sub module validates incoming agents. It is also responsible for maintaining trust between other AgentStation services.

## 3.3. Agents

Mobile Agents are entities that perform some job on behalf of its owner, can move around the network and get executed on different AgentStations. An agent gets loaded at the AgentStarter or the AgentStarter sub module at the Agent Station either locally or remotely from its jar file. When loaded at the AgentStation, the agent can move to any Agent Station for its execution.

The Service Agent (which extends the generic agent) contains a remote GUI [5] which enables a defined agent's GUI to be visible to the client. The GUI displayed at the client, is sent in response to a remote call on the agent, after a two way security hand-shake depending on the type of agent. If any field/component in the remote GUI [5] of the agent or AgentStation visible in the client module is displaying a value from where the Agent is actually executing, then that value will be determined by making a remote call to the agent itself. Service Agent can register itself with the LUS and any client can download its remote GUI [5].

If an agent wants to move to another AgentStation (either from an AgentStarter or from an AgentStation), it marshalls itself and calls a method on that AgentStation's proxy which allows the marshalled object to move to that AgentStation. The proxy then collects the marshalled object and the AgentStation is notified that its proxy, which was downloaded somewhere, has received an agent. The AgentStation then proceeds to collect that marshalled object from the proxy. The AgentStation then unmarshalls that agent and after carrying out some security checks, it allows its execution. If an agent is a 'Service' type (agents registered with LUS), then the AgentStation allows the agent to register with the LUS using the AgentStations discovery and registration interface.

The agent is moved as a marshalled object from one AgentStation to another (and also from AgentStarter to AgentStation). This marshalled object has the state of an agent saved inside it but no class definition. To unmarshall the object, the receiving AgentStation must know its class definition. The agent carries with itself a URL to the location where its class file is located. This class file is packed in a signed jar file and placed at an HTTP Server. When an AgentStation unmarshalls the agent, it also loads its class files after downloading the jar file from the HTTP Server

Transactions are used in the framework for providing atomic operations and to avoid being in an unstable state in case of failure. The agent movement process is made atomic with the use of transaction service provided by Jini. While moving from one AgentStation Service to another, the agent needs to un-register from the first AgentStation and register again after arriving at the second AgentStation. During this process, there is a chance of agent loss (if the agent fails to register itself at the destination AgentStation) or duplication (agent registered at two places). These situations are avoided by implementing an atomic move operation.

## 3.4. Client

The client module acts as a Jini client, thus is able to discover all the AgentStations and agents. The client module can be used to perform the monitoring of the AgentStation or the agents as both have registered their downloadable GUI's [5] in their proxies with the Lookup service.

The owner/creator uses the client module to download the GUI of the AgentStations to perform remote administration or to download the GUI of the agents to perform real time monitoring or to interact with the agent.

# 4. SECURITY

Agent based systems have an inherent need for security. The security requirements of such systems are twofold. Not only is there a need for protecting a host from a malicious agent, but it is also required that the mobile agent be protected from malevolent hosts. Security architecture for the DIAMOnDS system has been designed basing on the notion of trust i.e. services with the need of security don't cooperate with each other unless trust is established among them. Trust establishment is realized using the Public Key Infrastructure (PKI) [9]. Cooperating entities must exchange their certificates before providing any service to each other. Figure 2 shows that an AgentStation possesses a Private Key Store which contains the public and private keys of the administrator of this AgentStation. It also has a Trust Store which contains the exported certificates of trusted users who are the administrators of other AgentStations. When an agent moves, it carries the certificate of its owner/creator for authentication purposes as it would be acting on behalf of its owner/creator at the AgentStation and needs owner's/creator's credentials.

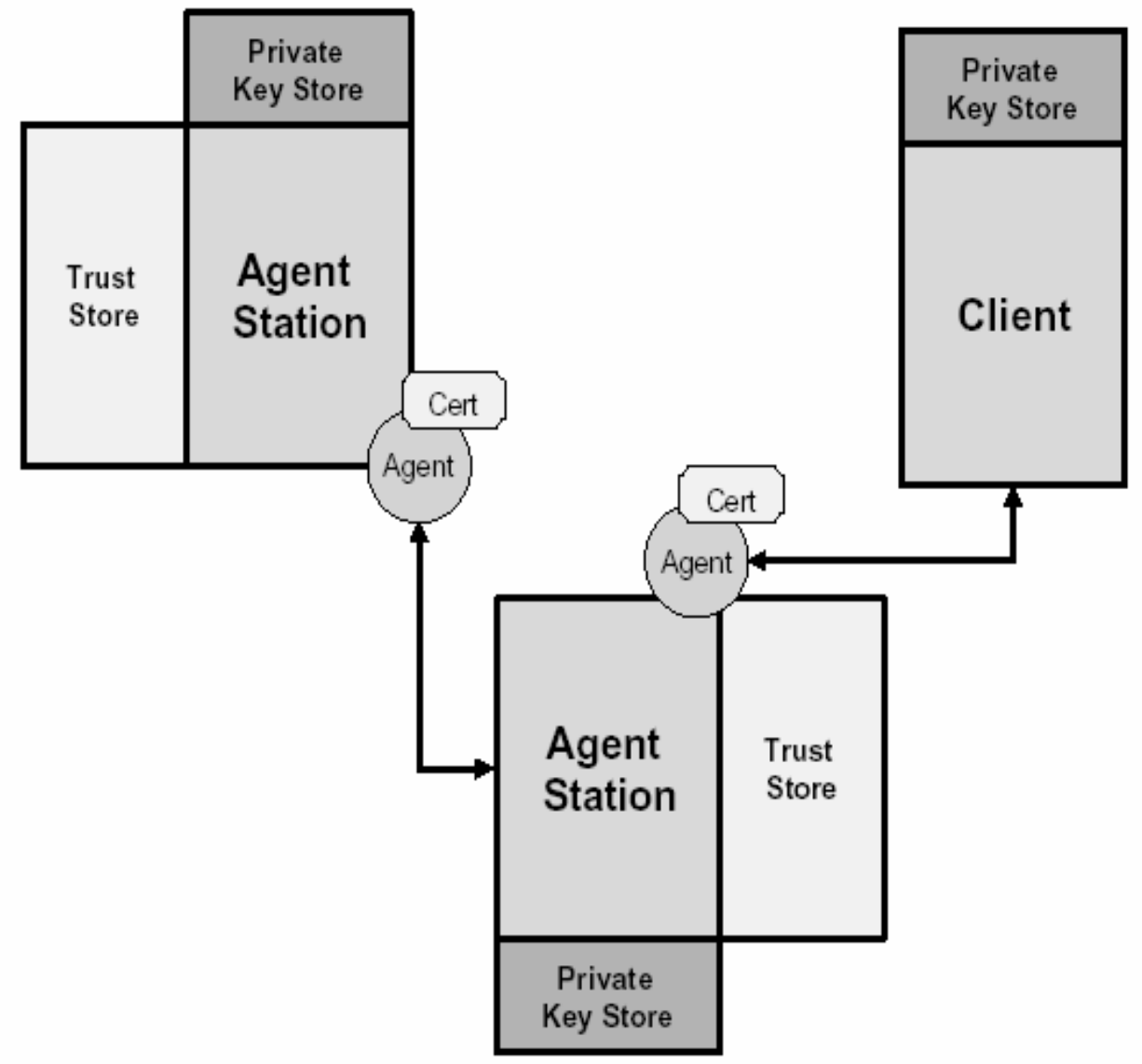

Figure 2: Security Mechanism of DIAMOnDS

At present, two basic security mechanisms have been implemented. The first involves trust management between the agent's owner/creator and the host (Agent Station). The owner of an agent signs its jar file with his private key. When a jar file is signed, the signer's certificate is also embedded in the jar file. Before moving an agent, the owner needs to export its certificate to the target AgentStation on which the owner/creator intends to execute the agent. Each AgentStation maintains a set of trusted certificates with itself. The owner of the AgentStation has the authority to accept or reject an arriving certificate. When an agent requests to move to an AgentStation, two checks are performed. First, the owner's certificate is checked to see if it is trusted or not. Second, the signature of the jar file providing the agent's codebase is verified against its contents, which determines whether or not there has been a change in the contents after they were signed. Such a change indicates malicious modification of classes not intended by the creator of the agent and is unacceptable. Failure of any of the above checks implies that the agent is not authenticated. If the checks are successful, the agent is allowed to execute at the AgentStation.

The second security measure is related to remote monitoring and administration of the agent. Considering the security aspects, there are two types of agent; one is open for monitoring and controlling while the second type is closed to everyone for this purpose other than the agent's creator/owner. An agent's GUI is not open to every client except the one who owns it, unless the owner allows it to be open to everyone. Downloading GUI of a protected agent involves a handshake between the agent and the client module. When a client needs to get an agent's GUI, it signs some randomly generated data with the agent's private key. This signed data is sent to the agent as part of the request to get the GUI. The agent tries to verify this signature with its owner's certificate. A successful verification means that the requestor is really the owner of the agent and the GUI is delivered to the client module for the owner to see it. A verification failure results in rejection of the request.

## 5. APPLICATION AGENTS

Three types of application agents have been developed on the framework. These agents extend the service agent i.e they act as Jini services.

### 5.1. File Access Agent

The File Access Agent serves the user by providing access to the file system of the AgentStation where it is executing. The owner/creator can be residing on a computer that is resource constrained or the user can be interested in some file store that is available on the remote AgentStation, then the user can send the File Access Agent to that AgentStation and by downloading the GUI of the File Access Agent in the client module, the owner/creator can read/write to the file system of the AgentStation. The results of the initial testing of the modules are mentioned later.

### 5.2. Database Access Agent

The AgentStation may provide the service of data store by establishing connections to various databases present on its LAN to the dynamic services that it can host. It keeps the updated information about these heterogeneous databases and the Database Access Agent is provided with all this information as it arrives on the AgentStation. The owner/creator can monitor the agent by downloading the agent's downloadable GUI in the client module. The owner/creator can interact with all the databases as if they are local databases, just sending the query and getting the results back.

### 5.3. Connectivity Test Agent

A Connectivity Test Agent has been developed that is able to travel between two or more AgentStations and keep a log of its arrival and departure time to test the validity of the code mobility achieved. Its purpose is only experimental and is used to test the validity of service availability and mobility issues. The results of the test conducted are described below.

## 6. TESTING

The framework has undergone considerable testing to verify the following:

- The basic agents are able to continue 'working' actively for long durations of time. This feature is desirable for specialized agents that will be used in data mining jobs.
- Read/write performance of File Access Agent is sufficiently comparable with Local File System, LAN and Network File System read/write mechanisms.
- The agents are not lost or duplicated during the movement process.

### 6.1. Mobility Test

The purpose of this test was to verify the above mentioned first two points. Moreover this test verifies that the mobility mechanism of this framework is stable enough to work in practical networks that are inherently unreliable. AgentStations were started at remote locations on WAN, one at NUST, Pakistan and the other at CERN, Switzerland. An agent was loaded at one of the AgentStation and given an itinerary path to keep moving between these two AgentStations, staying at one AgentStation for no more than 30 seconds. This agent was supposed to keep record of arrival and departure time at/from each AgentStation which could be retrieved later on. This agent continued its migration between the two AgentStations for nearly 12 hours without being lost or duplicated. The readings collected by this agent were also found to be correct and were not lost during agent's movement (the agent keeps the log within its migrating state and is not stored to any persistent storage during its movement and execution).

### 6.2. File Access Tests

These tests were aimed at measuring the performance of read / write operations through the File Access agent. Readings were taken by measuring the performance over LAN and WAN in comparison to carrying out these operations without agents. The following charts show the performance of the agent when in M Bytes/sec. The performance is slower than Network File System as the agents are built on top of RMI. (Tests were taken at both 100 Mbps and 10 Mbps Ethernet).

### 6.3. Improvement in File Access Rates

Originally the File Access agent transferred data through method calls back to the client and effectively all the data was transferred over RMI, thus seriously affecting data transfer rates. This mechanism was altered to transfer data over sockets (Java Sockets) and considerable improvement in data transfer rates was achieved. A comparison between data transfer rates over RMI and Sockets is given below in Figure 3 and 4.

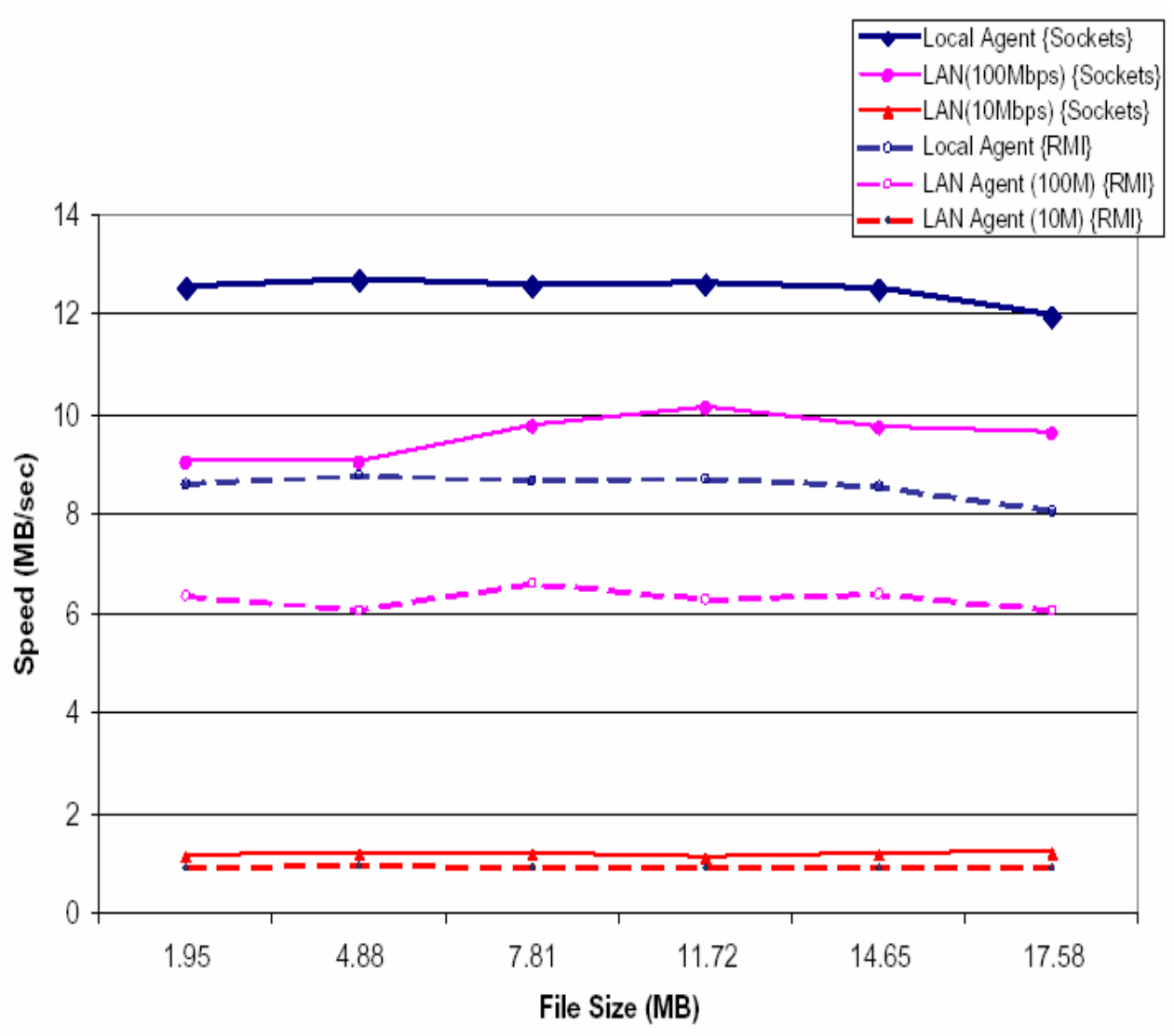

Figure 3: Read Performance of Agent – Comparison of Sockets & RMI Data Transfer

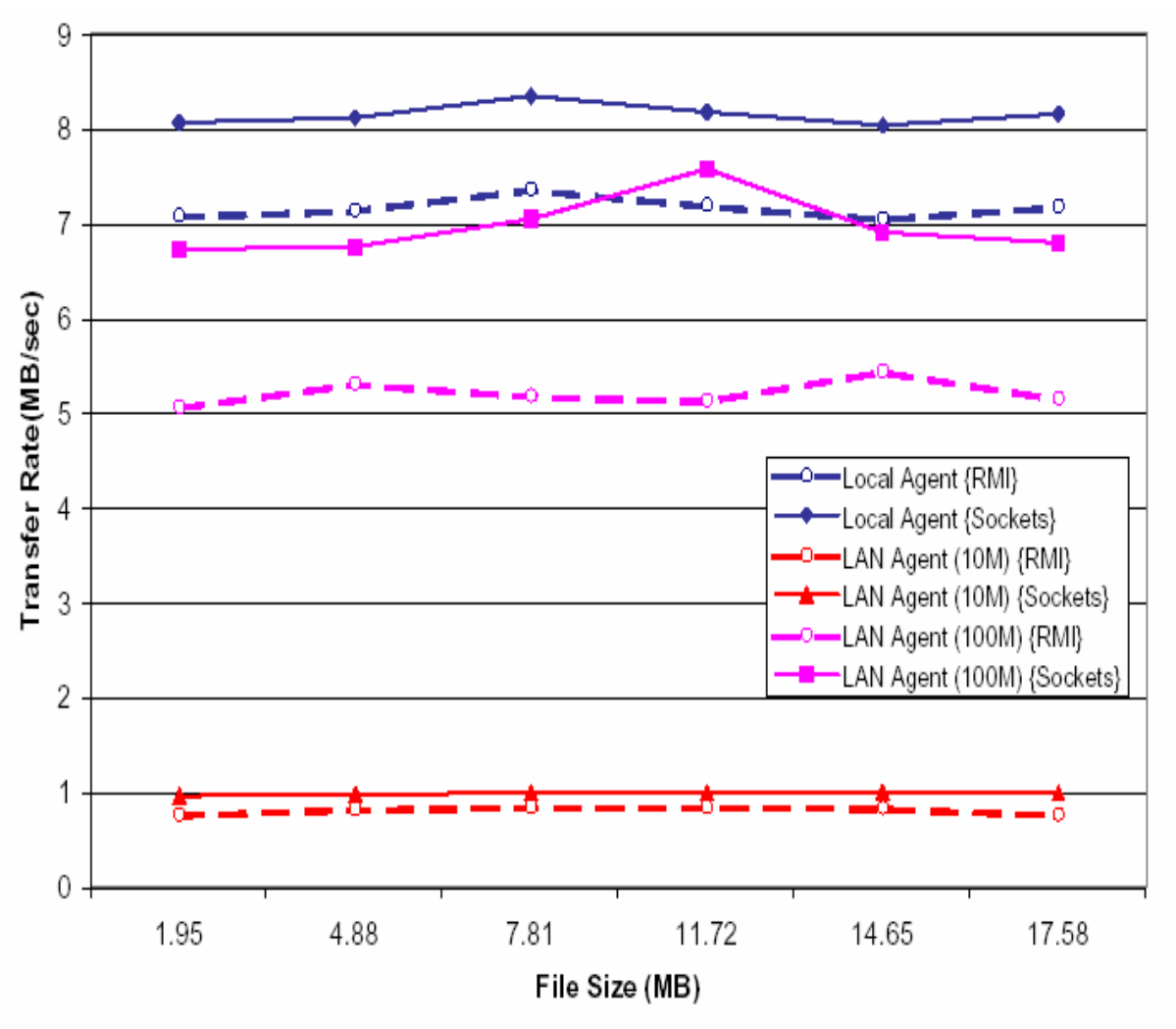

Figure 4: Write Performance of Agent – Comparison of Sockets & RMI Data Transfer

## 7. CONCLUSION

DIAMOnDS is an infrastructure for building mobile agents with downloadable GUIs. It is built completely on Java/Jini and besides allowing the end users to access their agents conventionally as in toolkits like IBM Aglets [10] and others, allows the agents to be published as services, which makes it possible to monitor them remotely and move around the network to optimally perform specific tasks for its end users. The innovative concept of mobile agents acting as mobile services in the Service Oriented Architecture brings forth the better of both worlds.

## 8. CURRENT STATUS AND FUTURE WORK

Development of two specific agents that are relevant to LHC experiments related data transfers is in progress.

One of the agents will be used to inform the agent's owner about the availability of a certain type of data in particular data stores. In this scenario, it is assumed that data generated through LHC experiments will be available through virtual data tags in data stores at different tier levels. The agents will be given criteria for search/requirement by their users and these agents will move to an AgentStation closer to the data store These agents will monitor the data stores for availability/generation of the query data type and on availability, it will inform its owner. Issues like timeout and close matches will be handled by the agent itself.

A basic data searching agent has been implemented that performs a query based search in a file system. This type of agent analyses the contents of a file and based on the query, assigns a relevance weight to the file. After searching through one or many file systems, it returns to the user and shows the result. The user can then download the files of his interest. Large scale data generated as a result of High Energy Physics [1] experiments is one possible area where complex data mining ability can be purposefully introduced in these agents.

The second type of agent related to LHC experiments is the 'Optimal Network Path Finding Agent' which will be used to determine optimal path for large scale data transfer between different tiers on wide area networks. These agents will collaborate to determine the best route for data connection in terms of bandwidth available, network load and other features that decide quality of service in wide area network connections. For the collection of such parameters, we intend to integrate this framework with MonaLisa [11].

Service UI specifications [5] allows to download the GUI of the services in some resource constrained environment. We intend to provide WAP access to agents and AgentStations. When in place, this will allow the user of a particular agent service to monitor the activities of his personalized agent through a WAP enabled device. A particular scenario for the user will be to determine, using his mobile phone, where his agents is, what is it doing and what portion of the job assigned to it has been completed. It will also allow him to control his agent remotely on a wireless device.

The security implemented in DIAMOnDS is embryonic at this stage. The final release of Jini 2.0 toolkit [12] has added many useful features like enhanced security, invocation constraints etc. which can prove useful in the enhancement of the existing security mechanisms.

## Acknowledgments

We would especially like to thank Dr H Farooq Ahmad, Specialist Engineer at Communication Technologies, Sendai Japan and Shahzad Khan, Assistant Professor at NUST Institute of Information Technology, Pakistan for their useful technical guidance and assistance during the project. We would also like to thank the administration and faculty members of National University of Sciences and Technology, Pakistan and California Institute of Technology, USA for their generous technical and moral support throughout this project.